\documentclass[sigconf]{acmart}

\usepackage{booktabs}
\usepackage{colortbl}
\usepackage{makecell}
\usepackage{multirow}
\usepackage{graphicx}

\usepackage{listings}
\usepackage[utf8]{inputenc}
\usepackage{xcolor}

\definecolor{codegreen}{rgb}{0,0.6,0}
\definecolor{codegray}{rgb}{0.5,0.5,0.5}
\definecolor{codepurple}{rgb}{0.58,0,0.82}
\definecolor{backcolour}{rgb}{0.95,0.95,0.92}

\lstdefinestyle{mystyle}{
  backgroundcolor=\color{backcolour},   commentstyle=\color{codegreen},
  keywordstyle=\color{magenta},
  numberstyle=\tiny\color{codegray},
  stringstyle=\color{codepurple},
  basicstyle=\ttfamily\scriptsize,
  breakatwhitespace=false,         
  breaklines=true,                 
  captionpos=b,                    
  keepspaces=true,                 
  showspaces=false,                
  showstringspaces=false,
  showtabs=false                 
}
\AtBeginDocument{%
  \providecommand\BibTeX{{%
    \normalfont B\kern-0.5em{\scshape i\kern-0.25em b}\kern-0.8em\TeX}}}

\copyrightyear{2021} 
\acmYear{2021} 
\setcopyright{acmcopyright}\acmConference[CIKM '21]{Proceedings of the 30th ACM International Conference on Information and Knowledge Management}{November 1--5, 2021}{Virtual Event, QLD, Australia}
\acmBooktitle{Proceedings of the 30th ACM International Conference on Information and Knowledge Management (CIKM '21), November 1--5, 2021, Virtual Event, QLD, Australia}
\acmPrice{15.00}
\acmDOI{10.1145/3459637.3482191}
\acmISBN{978-1-4503-8446-9/21/11}

\usepackage{float} 
\usepackage{physics}

\usepackage{pgfplots}
\pgfplotsset{compat=1.9}
\usepackage{pgfplotstable}
\usepgfplotslibrary{groupplots}

\usepackage{graphics}      

\usepackage{algorithm}
\usepackage{algorithmic}

\usepackage{pgfplots}
\pgfplotsset{compat=1.8}
\usepackage{pgfplotstable}
\usepgfplotslibrary{groupplots}

\usepackage{fancyhdr}

\begin{document}
\fancyhead{} 
\title{Storing Multi-model Data in RDBMSs based on Reinforcement Learning}

\author{Gongsheng Yuan}
\email{gongsheng.yuan@helsinki.fi}
\affiliation{%
  \institution{University of Helsinki \& Renmin University of China}
  \city{Helsinki}
  \country{Finland}
}

\author{Jiaheng Lu, Shuxun Zhang, Zhengtong Yan}
\email{{jiaheng.lu, shuxun.zhang, zhengtong.yan}@helsinki.fi}
\affiliation{%
  \institution{University of Helsinki}
  \city{Helsinki}
  \country{Finland}}





\begin{abstract}


How to manage various data in a unified way is a significant research topic in the field of databases. To address this problem, researchers have proposed multi-model databases to support multiple data models in a uniform platform with a single unified query language. However, since relational databases are predominant in the current market, it is expensive to replace them with others. Besides, due to the theories and technologies of RDBMSs having been enhanced over decades, it is hard to use few years to develop a multi-model database that can be compared with existing RDBMSs in handling security, query optimization, transaction management, etc. In this paper, we reconsider employing relational databases to store and query multi-model data. Unfortunately, the mismatch between the complexity of multi-model data structure and the simplicity of flat relational tables makes this difficult. Against this challenge, we utilize the reinforcement learning (RL) method to learn a relational schema by interacting with an RDBMS. Instead of using the classic Q-learning algorithm, we propose a variant Q-learning algorithm, called \textit{Double Q-tables}, to reduce the dimension of the original Q-table and improve learning efficiency. Experimental results show that our approach could learn a relational schema outperforming the existing multi-model storage schema in terms of query time and space consumption.

\end{abstract}







\maketitle

\section{Introduction}

With the development of technology, different users may like to use different devices to collect relevant information from various perspectives for better describing or analyzing an identical phenomenon. For convenience, these devices with their supporting software would store the collected data in the way they like (e.g., using relational tables to preserve structured tabular data, using JSON document to record unstructured object-like data, and using RDF graph to store highly linked referential data), which is inevitable to cause data variety. Although choosing different databases to manage different data models is an ordinary operation, this would result in operational friction, latency, data inconsistency, etc., when using multiple databases in the same project.

\begin{figure}
\centering
\includegraphics[height=3.5cm,width=7.3cm]{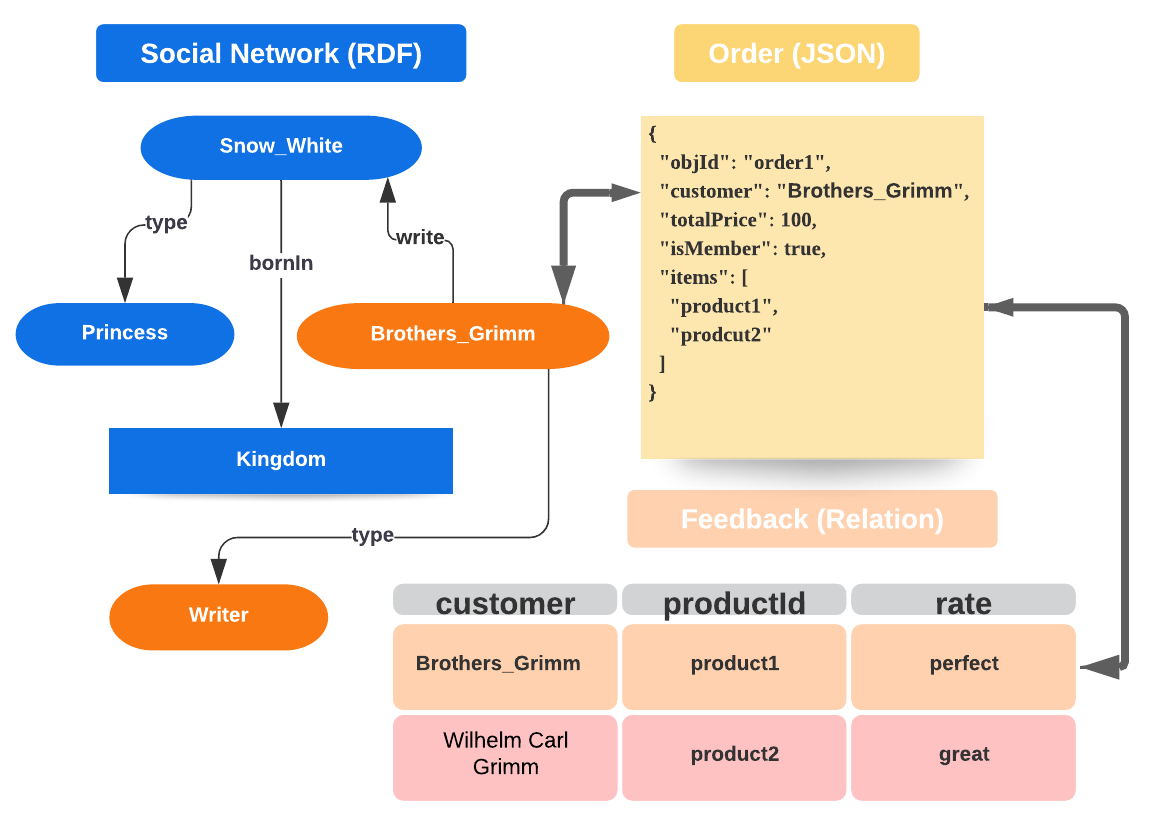}
\caption{An Example of Multi-model Data.}
\label{fig:multimodel}
\end{figure}

Against this issue, researchers propose a multi-model database concept that manages multi-model data (see Figure~\ref{fig:multimodel}) in a unified system \cite{lu2019multi, lu2018multi}. However, since researchers have spent decades strengthening RDBMS theories and technologies, it is not easy to use few years to develop a multi-model database to effectively and efficiently handle security, query optimization, transaction management, etc. Besides, since relational databases are still predominating the current market and storing mass legacy data, it is not easy to replace RDBMSs with multi-model databases at a low cost. Therefore, it fuels more and more interest to reconsider loading and processing multi-model data within RDBMSs.

JSON, as the representative of semi-structured data, is proposed as a hierarchical, schema-less, and dynamic data interchange format on the web. Each JSON object comprises a set of structured key-value pairs, where key denotes attribute name, value is the attribute value. As for RDF data, it is a collection of statements where each statement is defined as \emph{subject-predicate-object (s,p,o)} meaning that a subject $s$ has a predicate (property) $p$ with value $o$. For storing them in RDBMSs, one straightforward method is to map the JSON document into several three-column tables (each table consists of object ID, key name, and value) \cite{chasseur2013enabling} and parse the RDF graph into a series of triples stored in a three-column table \cite{mcbride2002jena}. Unfortunately, this method not only involves many self-joins, but it ignores the relationships among this multi-model data.


After reviewing the literature, we found there was still a blank for this research. The current works only focus on storing semi-structured document in RDBMSs \cite{deutsch1999storingSTORED,chasseur2013enabling} or storing graph data in RDBMSs \cite{zheng2020efficient, sun2015sqlgraph}. Due to the mismatch between the complexity of multi-model data structure (as shown in Figure~\ref{fig:multimodel}) and the simplicity of flat relational tables, we think it is a challenge to design a good schema for storing multi-model data in an RDBMS.

In this research, we attempt to generate a good relational schema based on the RL method to store and query multi-model data consisting of relational data, JSON documents, and RDF data in RDBMSs. It is not easy to employ the RL method to learn a relational schema for multi-model data. We need to define its states, actions, rewards, etc. And we propose a Double Q-tables method as a learning algorithm to help the agent choose actions, which extremely reduce the dimensions of the original Q-table's action columns and improve learning efficiency.

The motivation for using RL is that RL is an optimization problem, and we could use Markov Decision Process to model the process of relational schema generation.
Given an initial state (schema), RL could work with a dynamic environment (RDBMSs) and find the best sequence of actions (joins) that will generate the optimal outcome (relational schema collecting the most reward), where the most reward means the final generated schema have the minimum query time for a set of queries on the given workload. Specifically, we use the Q-learning algorithm to take in state observations (schemas) as the inputs and map them to actions (joins) as the outputs. In RL, this mapping is called the policy, which decides which action to take for collecting the most reward over time rather than a short-term benefit. It is just like supervised learning.

The rest of this paper is organized as follows. Section 2 introduces our approach. Section 3 shows the preliminary experimental results. Finally, the last section summarizes this paper.

\begin{figure}
\centering
\includegraphics[height=3.8cm]{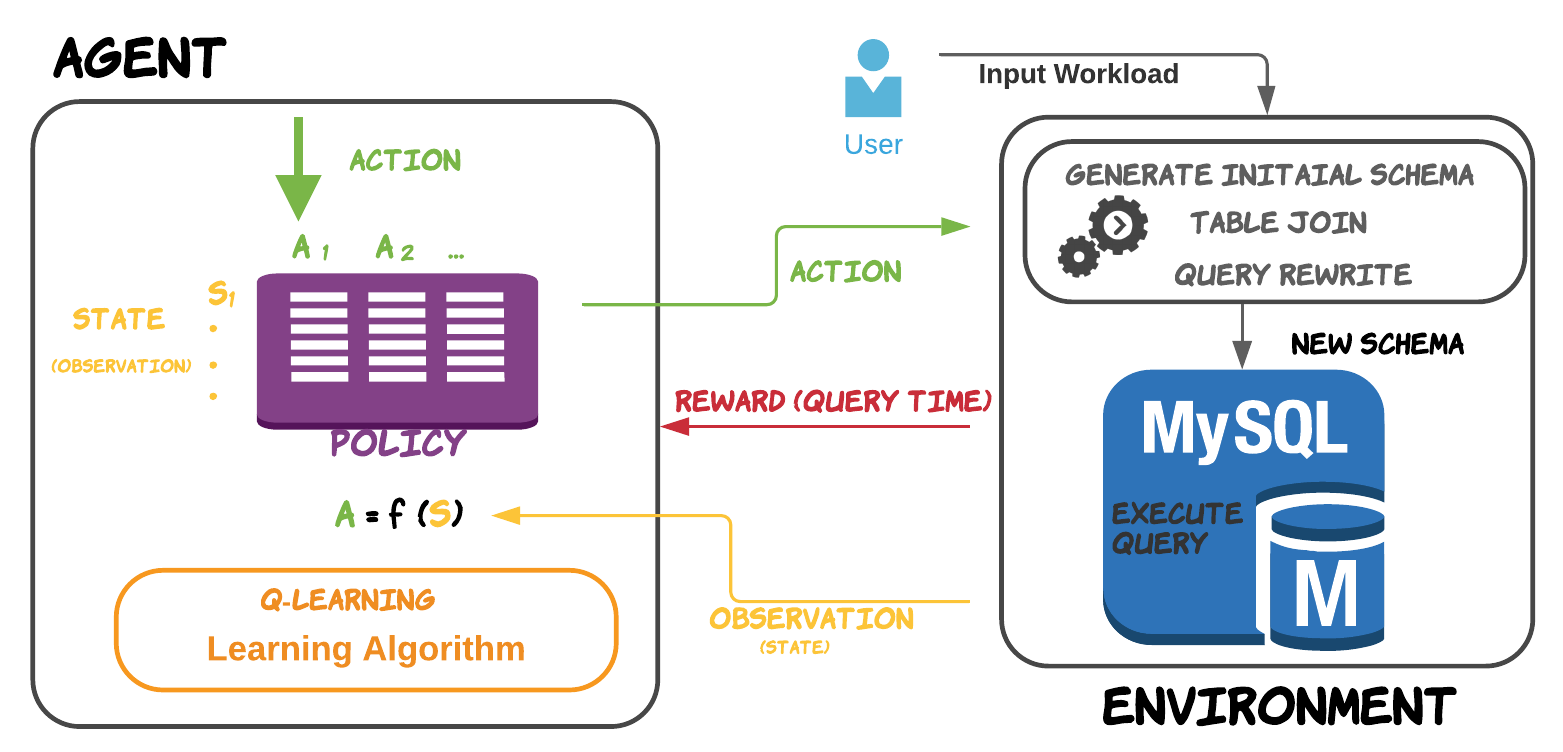}
\caption{The Framework of Transforming Multi-model Data into Relational Table based on RL}
\label{fig:framework}
\end{figure}

\section{The Proposed Approach}

\subsection{The Overview of Framework}
Since RL allows an agent to explore, interact with, and learn from the environment to maximize the cumulative reward, we propose a transforming multi-model data into relational tuples framework based on reinforcement Learning to store and query multi-model data in RDBMSs (see Figure~\ref{fig:framework}). In this framework, we first generate an initial schema to be the initial state. Next, according to the current schema (state) and policy, we choose an action to generate a new (next) schema (state). Then, we rewrite the workload query statements to adapt the new schema and perform the rewritten query statements on the generated schema in MySQL databases. After that, the MySQL database returns the query time. And we regard the increase in negative query time compare to the previous state as a reward. Finally, the agent updates the Q-table based on the returned reward and observation (state). And it re-chooses an action to explore the potential relational schema or collect the most rewards that we already know about until the agent has tried all the actions in this episode (or the episode has reached the maximum number of iterations). After running the episode as many times as you want and collecting the generated states and their query time in this process, we could obtain a good relational schema for this multi-model data workload.



To formalize the problem of generating a relational schema for multi-model data as a reinforcement learning problem, we need to figure out input, goal, reward, policy, and observation. Next, we will introduce them separately.

\begin{table}[!htbp]
\centering
\setlength{\extrarowheight}{0pt}
\addtolength{\extrarowheight}{\aboverulesep}
\addtolength{\extrarowheight}{\belowrulesep}
\setlength{\aboverulesep}{0pt}
\setlength{\belowrulesep}{0pt}
\caption{ArrayStringTable}
\label{tab:arrayTableFigure}

\begin{tabular}{p{1.7cm}<{\centering}|p{1.7cm}<{\centering}|p{1.7cm}<{\centering}|p{1.7cm}<{\centering}}
\midrule
\rowcolor{lightgray!20}
\textbf{objId} & \textbf{key} & \textbf{index} & \textbf{valStr} \\ 
\toprule
1 & items & 0 & product1 \\ 
\hline
1 & items & 1 & product2  \\ 
\bottomrule
\end{tabular}
\end{table}

\subsection{Initial State}
This framework uses a fully decomposed storage model (DSM) \cite{copeland1985decomposition} to preserve multi-model data as the initial schema. Besides, we adopt the model-based method \cite{florescu1999storing} to design a fixed schema for preserving JSON arrays. In detail, if the number of unique keys in the JSON document is ($n$ + 1), we decompose a JSON document into $n$ two-column tables. Please note that we do not take JSON object id into account. We will use it to distinguish to which object these keys belong. For each table, the first column contains the JSON object id, and the second column holds the values for the corresponding key. Moreover, we use these keys as two-column table names. The table in Table~\ref{tab:arrayTableFigure} is called ArrayStringTable, which is designed to store array elements that have string type value. Similarly, we could also define ArrayNumberTable and ArrayBoolTable.


For the relational data, we split each table into multiple little tables whose amounts are equal to the number of unique attributes except the table keys. For each little table, the first several columns contain the original tuple keys, and the following column holds the values for the corresponding attribute (This attribute is also the name of this little table).

For the RDF graph, we break a triples table into multiple two-column tables whose amounts are equal to the number of unique predicates in this dataset. Within each table, the first column holds the subjects having the specific predicate, and its corresponding object values are preserved in the second column. Here, we use these predicates as two-column table names.

The benefits of using DSM are: (1) it could handle a dynamic dataset. If there is a new appearing attribute, we could add a new table to preserve it, and there is no need to change the current schema; (2) it could reduce the complexity of actions. In the next part, we will give the definition of actions.

\subsection{Action}

First, we collect and count all the keys (JSON), predicates (RDF), and attributes (relation) in the multi-model data. Next, we map each JSON key to an integer id from 1 to $n$, map each predicate to an integer id from ($threshold_1 + 1$) to ($threshold_1 + m$), and map each attribute to an integer id from ($threshold_2 + 1$) to ($threshold_2 + p$) where $n$ is the total number of keys, $m$ is the number of predicates, $p$ is the number of attributes, and $threshold_1$ is a number that is used to distinguish keys from predicates (i.e., all the ids of JSON keys are less than $threshold_1$ and all the predicate ids are greater than $threshold_1$ and less than $thredhold_2$). Similarly, we set $threshold_2$ to distinguish attributes from keys and predicates. Finally, we make each id represent a \textit{\textbf{join}} action. This means when the agent chooses an id, we will use its representative table to do joining. These ids (keys, predicates, and attributes) form a set of actions, denoted as $A$. For example, we could get $A$ = \{1, 2, 3, 4, 11, 12, 13, 21\} from Figure~\ref{fig:multimodel}, where \{$1:customer$\}, \{$2:totalPrice$\}, \{$3:isMember$\}, \{$11:type$\}, \{$12:bornIn$\}, \{$13:write$\}, \{$21:rate$\}, $threshold_1 = 10$, and $threshold_2 = 20$. In general, we are used to regarding a ($table_1$, $table_2$) pair as an action to indicate which two tables to do joining. But, this is easy to form a large action space.
The benefit of our action definition is to reduce the size of action space into $O(q)$ where $q = n + m + p$.

\subsection{State}

Since we have known all the little table names and their corresponding mapping ids, we could use them to represent the states (i.e., relational schemas). For example, we could represent initial state $s_0$ as a string "1 0 2 0 3 0 11 0 12 0 13 0 21". We use the reserved word "0" as an interval bit between different tables in this expression. To clearly express the relational schema, we put these ids in numerical order even inside a table. For example, in the state $s_1$ ("1 3 0 2 0 11 0 12 0 13 0 21"), it has a table [1, 3] where the attribute 1 and attribute 3 are arranged in ascending order. Please note that we use a dictionary $D$ instead of a string to represent a relational schema, although they have the same meaning. For example, the relational schema of state $s_0$ is $D_0$ = \{1:[1], 2:[2], 3:[3], 4:[4], 11:[11], 12:[12], 13:[13], 21:[21]\}. In this dictionary, each key represents a table id, and its corresponding value is its table attributes consisting of keys (JSON), predicates (RDF), or attributes (relation). And all of the keys in $D_0$ represent the current existing tables at the state $s_0$.


\begin{algorithm}[t]
\begin{algorithmic}[1]
    \renewcommand{\algorithmicrequire}{\textbf{Input:}}
    \renewcommand{\algorithmicensure}{\textbf{Output:}}
    \REQUIRE The multi-model data and queries
    \ENSURE  A good relational schema
    \STATE  Initialize $QT_{a}$ and $QT_{join}$
    \STATE  Generate initial relational schema $D_0$ and initial state $s_0$
    \FOR {each episode} 
         \STATE $D$ = $D_0$, $s$ = $s_0$
         \STATE Initialize action space $A$
         \WHILE{True}
         
            \STATE Choose a action $a$ at state $s$ by $QT_{a}$ ($\epsilon$-greedy)
         
            \STATE Remove the action $a$ from the action space $A$
         
            \STATE Choose a table id at state $a$ by $QT_{join}$ ($\epsilon$-greedy)
         
             \STATE Execute join, perform queries, and observe $r$, $s'$
         
            \STATE Update $QT_{a}$ and $QT_{join}$
            
            
            
         
            \STATE $s$ $\gets$ $s'$
            
            \STATE until $A$ is empty
         
         \ENDWHILE
         
    \ENDFOR
\end{algorithmic}
\caption{Generating a Relational Schema based on RL (GRSRL)}
\label{alg:approach}
\end{algorithm}

         
         
         
         
         
            
            
            
         
            
         
         

\subsection{Policy}

In the reinforcement learning framework, the agent knows nothing about the environment. It just knows that it can take in a state and one possible action from that state and get its new state and rewards back from the environment after taking that action. Since there is a finite number of states and actions in our problem, we adopt the classic Q-learning algorithm \cite{watkins1989learning} for action selection, whose core idea is to generate a Q-table ($QT_{a}$) to store state-action values.


Since we define actions as join operations, we could not make it work just with one Q-table and the current state, except that we only do self-joins.
This is because we have no idea which table should be chosen from the current schema to do joining with the table to which the action $a_i$ corresponds.


To address this problem, we have defined another Q-table $QT_{join}$. It is a $(q \times q)$-dimensional table whose rows (states) are defined by $A$, columns (actions) are defined by table ids, meaning that when $QT_{a}$ chooses an action $a_i$, the $QT_{join}$ will decide which table (selected from the current schema) should be chosen to do joining with $a_i$ ($QT_{a}$) at the state $a_i$ ($QT_{join}$). For example, in the state $s_0$ of $QT_{a}$, $QT_{a}$ chooses an action 3 and removes the action 3 from its action space. Then if $QT_{join}$ chooses a table id (tID) 1 (action of $QT_{join}$) from the current schema at state 3 (state of $QT_{join}$), the new state of $QT_{a}$ will be state $s_1$, and the $D_1$ = \{1:[1,3], 2:[2], 4:[4], 11:[11], 12:[12], 13:[13], 21:[21]\}. The agent updates the two Q-table values until the action space $A$ of $QT_{a}$ is empty or the episode reaches the maximum number of iterations. Besides, we set a sematic constraint pool (including like key and foreign key constraints) to determine whether they do joining.






\subsection{Reward and Goal}

The reward is an instantaneous benefit of being in a specific state after the agent executing an action. In our problem, first, we obtain the negative value of query time gotten by the MySQL database performing a set of workload queries on the current relational schema, denoted as $-t_{n}$. Then, we define the reward as the reduction of this value compared to the previous negative query time, i.e., ($t_{p}$ - $t_{n}$). This is because our goal is to let our agent automatically learn a relational schema having the minimum query time for a set of queries on the given workload by interacting with the MySQL database environment.

Based on the above concepts, we propose Algorithm~\ref{alg:approach} to describe our learning process.

\section{Experiment}

We use the Person dataset\footnote{https://www2.helsinki.fi/en/researchgroups/unified-database-management-systems-udbms/datasets/person-dataset} to compare our approach's performance with the multi-model database ArangoDB's. After conducting the data clean, we list its statistics in Table~\ref{tab:dataset}.

\begin{table}[!htbp]
\caption{The number of triples/objects in the data}
\centering
\label{tab:dataset}
\begin{tabular}{p{2cm}<{\centering}|p{2cm}<{\centering}|p{2cm}<{\centering}}
\hline
            &  RDF    &   JSON \\ \hline
Person      &   4 471 790 &  153 134 \\\hline
\end{tabular}
\end{table}

\begin{table}[!ht]
\footnotesize
\small
\centering
\caption{Queries employed in the experiments}
\label{tab:queries}
\begin{tabular}{ll}
\toprule
\multicolumn{1}{l}{ID}  \quad  &   \multicolumn{1}{l}{Queries}     \\ \midrule
$Q_{1}$  & Return Doris\_Brougham's pageid \\
$Q_{2}$  & Return all the values of subject about Tor\_Ahlsand \\
$Q_{3}$  & Return birtthDate, activeYearsStartYear, \\
         & and activeYearsEndYear of Heath\_Ledger \\
$Q_{4}$  & Return original and title when pageid = 8484745 \\ 
$Q_{5}$  & Return birthYear and ns of Sadako\_Sasaki  \\
\bottomrule
\end{tabular}
\end{table}

As shown in Table~\ref{tab:queries}, we employ these queries in our experiments, where each query includes one or two data models. For each $Q_i$, we perform the corresponding AQL in ArangoDB databases to get its query time. Besides, we perform all queries three times and use their median value in our experiments.

The experiments are implemented in Python and run on a laptop PC with an Intel Core i7 processor of 3.19 GHz and 16GB memory. The version of MySQL is 5.7.30, ArangoDB is 3.7.1. And we set the maximum number of iteration as 100.


\begin{table}[!htbp]
\centering
\footnotesize
\setlength{\extrarowheight}{0pt}
\addtolength{\extrarowheight}{\aboverulesep}
\addtolength{\extrarowheight}{\belowrulesep}
\setlength{\aboverulesep}{0pt}
\setlength{\belowrulesep}{0pt}
\caption{Query Time (s)}
\label{tab:TotalTime}

\begin{tabular}{p{1.7cm}<{\centering}|p{0.7cm}<{\centering}|p{0.7cm}<{\centering}|p{0.7cm}<{\centering}|p{0.7cm}<{\centering}|p{0.7cm}<{\centering}|p{0.7cm}<{\centering}}
\midrule
\rowcolor{lightgray!20}
\textbf{Approaches} & \textbf{$Q_1$} & \textbf{$Q_2$} & \textbf{$Q_3$} & \textbf{$Q_4$} & \textbf{$Q_5$} & \textbf{Total}\\ 
\toprule
GRSRL & 0.265 & 1.789 & 0.106 & 0.201 & 0.373 & 2.733 \\ 
\hline
ArangoDB & 0.062 & 1.475 & 5.591 & 0.057 & 7.78 & 14.965  \\ 
\bottomrule
\end{tabular}
\end{table}

Table~\ref{tab:TotalTime} presents the executing time of queries on the relational schema generated by GRSRL and on ArangoDB. Due to ArangoDB storing data in document format, we can see the query time of ArangoDB is less than our schema when queries ($Q_1$, $Q_4$) only involve JSON. For the RDF queries(e.g., $Q_3$), our schema sometimes is better than Arangodb since the triples store requires many self-joins. Concerning the multi-model query of $Q_5$, our schema consisting of a mix of binary tables and property tables is better than ArangoBD.

\begin{figure}[!tbp]
\centering
\tiny
\begin{tikzpicture}
 \pgfplotstableread{dataPerson.csv}{\loadeddataone}
 \begin{groupplot}[
     group style={group size=1 by 1, horizontal sep=2em, vertical sep=7em},
     height = 2.55cm,
     width = 9cm,
    ]
    
    
     \nextgroupplot[
        xmin = 0, 
        ymax = 3.5,
		xlabel= episodes,
		ylabel=The total query time,
		ymin = 2,
		ylabel style={at={(-0.09,0.5)},anchor=north},
		xtick = {0, 10, 20, 30, 40, 50, 60, 70, 80, 90, 100},
		xticklabels = {0, 10, 20, 30, 40, 50, 60, 70, 80, 90, 100},
		]
		
        \addplot[color=red,mark=diamond,thick] table [x = number, y = value] {\loadeddataone};
        

    \end{groupplot}

\end{tikzpicture}
\caption{The changes in query time as the increase of episodes}
\label{fig:convergence}
\end{figure}
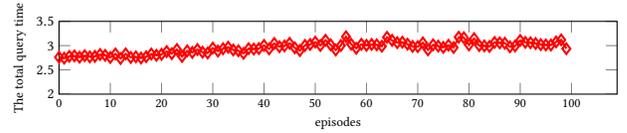


We set $\epsilon\_greedy = 0.1$ to try its best to explore different schemas. Then we use the optimal schema's query time of each episode to get Figure~\ref{fig:convergence} that manifests the trend of total query time as the increase of episodes. Of course, we could set a big number for $\epsilon\_greedy$ to make it converge fast.




	

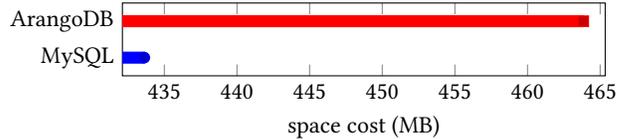
\begin{figure}[!tbp]
\centering
\begin{tikzpicture}
\begin{axis}[
    /pgf/number format/1000 sep={},
	enlargelimits=0.05,
	width=8cm, height=2.55cm, enlarge y limits=0.5,
	legend style={at={(0.5,-0.1)},
	    anchor=north,legend columns=-1},
	xlabel= {space cost (MB)},
	ytick={1,2},
	y tick style={draw=none},
	yticklabels={MySQL,ArangoDB},
	every axis plot/.append style={
          xbar,
          bar width=.3,
          bar shift=0pt,
          fill
        }
]
\addplot coordinates {(433.625,1)};
\addplot coordinates {(463.837,2)};
\end{axis}
\end{tikzpicture}
\caption{The cost of storage spaces on MySQL and Arangodb}
\label{fig:space}
\end{figure}

Figure~\ref{fig:space} shows the cost of storage spaces on MySQL and Arangodb, which denotes the cost of storage spaces of our schema on MySQL is less than Arangodb's.

\section{CONCLUSION}

In this paper, we employ a reinforcement learning method to propose a framework that could automatically learn a relational schema having the minimum query time for a set of queries on the given multi-model data workload by interacting with the MySQL database environment. Besides, we define the state, action, reward, etc., to support this RL-based relational schema generation framework. Especially, the definition of actions extremely reduces the dimension of the Q-table. The introduction of the Double Q-tables idea guarantees this framework to work successfully. Finally, the experiments show that our approach, GRSRL, could generate a good relational schema. And it offers the possibility of storing multi-model data in the RDBMSs while having a good performance.



\begin{acks}
The work is partially supported by the China Scholarship Council and the Academy of Finland project (No. 310321). We would also like to thank all the reviewers for their valuable comments and helpful suggestions.
\end{acks}


\bibliographystyle{ACM-Reference-Format}

\bibliography{sample-sigplan.bbl}


\end{document}